\documentclass[preprint,showpacs,preprintnumbers,amsmath,amssymb]{revtex4}

\usepackage{graphicx}
\usepackage{dcolumn}
\usepackage{bm}
\begin{document}
\title{Damping of Nodal Fermions Caused by a Dissipative Mode
}
\author{D. Schmeltzer}

\affiliation{Physics Department \\ City College of the City
University of New York
\\ New York, New York 10031}

\date{\today}
\begin{abstract}
Using a $d_{x^2 - y^2}$ superconductor in 2+1 dimensions we show
that the Nambu Goldstone fluctuations are replaced by dissipative
excitations.  We find that the nodal quasi-particles damping is
caused by the strong dissipative excitations near the nodal
points.  As a result we find that the scattering rates are linear
in frequency and not cubic as predicted in the literature for the
``d'' wave superconductors.  Our results explain the recent angle
resolved photoemission spectroscopy and optical conductivity in
the BSCCO high $T_c$ compounds.
\end{abstract}

\maketitle
The recent angle-resolved photoemission spectroscopy (ARPES) [1]
and optical conductivity [2] studies of the superconducting high
$T_c$ compound BSCCO show scattering rates which are linear in
temperature [1,2,3].  Moving away from the nodal direction the
scattering rates appear to level of and become temperature
independent.  These scattering rates are in strong contrast to the
$\omega^3$ scattering suggested for the d-wave like pairing [4].

Our results are expected to hold for any $d_{x^2 - y^2}$
superconductor in 2+1 dimension.  The fact that our predictions
are only observed in BSCCO suggest that the YBCO superconductors
have additional d-wave order parameters ``$id_{xy}$'', ``s'' or
have a large hopping in the ``c'' direction.  In both situations
the dissipative behavior caused by massless Dirac fermions in 2+1
is absent!

Quasi-particles properties of nodal fermions in $d_{x^2 - y^2}$
superconductors and 2+1 dimensions are investigated.  We find that
the Nambu-Goldstone fluctuations or the massive plasma mode caused
by the Coulomb interaction is replaced by a dissipative collective
mode - a gauge invariant field.  This result follows from the fact
that the Nodal-Dirac [5-9] fermions which couples to a gauge field
in 2+1 dimension induces critical Q.E.D. and not regular photons
[5-10]. We find that this dissipative mode causes the
quasi-particle self energy in the vicinity of the nodal points to
have scattering rate which are linear in frequency [1-3]. Moving
away from the nodal point one finds that the scattering rate is
reduced.  When the d-wave order parameter has additional
components, ``$id_{xy}$'', ``s'', $a'$ or single electrons hopping
occurs in the ``c'' direction the scattering rate of the
quasi-particles is suppressed.

In order to understand the origin of the dissipative mode we
follow S. Weinberg [10] and replace the fermion operator
$C_\sigma(x)$ by a neutral fermion field
$\tilde{C}_\sigma(x)=exp(- i \alpha(x)/2)C_\sigma(x)$ where
``$\alpha$'' is the Nambu-Goldstone phase.  In the presence of an
electromagnetic field $a_\mu^{ext}$ one finds [10] that the
electromagnetic response is represented in terms of the gauge
invariant fields $a_\mu = (\partial_\mu \alpha - 2 a_\mu^{ext})$.
For any superconductor the electromagnetic response contains two
parts, the diamagnetic contribution given by the vector potential
square, $\frac{1}{2} \rho (\vec{\partial} \alpha - 2
\vec{a}^{ext})^2$ ($\rho$ is the electron density) and the
paramagnetic response which is obtained by second order
perturbation theory.

For an ``s'' wave superconductor the density-density response
function is proportional to the superconducting density [12].  The
current-current response function represents the parametric
polarization, $\Gamma_2^{(i,j)}$ which at $T=0$ and $q \rightarrow
0$ vanishes, $\Gamma_2^{(i,j)} = 0$!  Therefore the diamagnetic
term $\frac{1}{2}\rho(\vec{\partial}\alpha - 2 \vec{a}^{ext})^2
$is not normalized.  As a result the Lagrangian takes the form;
$L^{sc}\sim \frac{1}{2}\rho\left[ \frac{1}{v^2}(\partial_t \alpha
- 2 a_0^{ext})^2 - (\vec{\partial}\alpha - 2 \vec{a}^{ext})^2
\right]$.  The Lagrangian $L^{sc}$ gives rise to the low energy
gapless mode - named Goldstone mode.  The presence of the Coulomb
interaction makes the Goldstone mode massive.  This can be seen
following the method presented in ref 12.  The Coulomb interaction
is replaced by a Hubbard-Stratonovich field $\tilde{a}_0$, which
play the role of the temporal gauge field. $L^{sc+c} \sim
\frac{1}{2}\rho \left[ \frac{1}{v^2}(\partial_t\alpha - 2
\tilde{a}_0)^2 - (\vec{\partial} \alpha)^2 \right] + \frac{1}{e^2
8 \pi}|\vec{q}|^2 \tilde{a}_0 \tilde{a}_0$ ``e'' is the electric
charge and $|\vec{q}|$ is the momentum. Integration of the
$\tilde{a}_0$ field changes the Goldstone mode into a massive
mode!

The $d_{x^2 - y^2}$ case in 2+1 is different!  In this case the
paramagnetic response is not zero and gives rise to a dissipative
mode which does'nt become massive in the presence of the Coulomb
interaction. The difference is caused by the polarization diagram
(see ref. 10, 12).  Instead of $\Gamma_2^{(0,0)} \sim \rho_s$ and
$\Gamma_2^{(i,j)} = 0$ we have a branch-cut,
$\Gamma_2^{(\mu,\nu)}(q) = \frac{1}{8\sqrt{-q^2}}(-q^2 g^{\mu\nu}
+ q^\mu q^\nu)$.  As a result we have; $L^{d-wave} \sim
\frac{1}{2}(\partial_\mu \alpha - 2 a_\mu^{ext})_{\vec{q}, \omega}
\Gamma_2^{(\mu,\nu)}(\vec{q},\omega)(\partial_\nu \alpha - 2
a_\nu^{ext})_{- \vec{q},- \omega} -
\frac{1}{2}\rho(\vec{\partial}\alpha -2 \vec{a}^{ext})^2$.  The
most important difference in $L^{d-wave}$ is the fact that the
term $(\partial_t \alpha - 2 a_0^{ext})^2$ is absent, instead we
have the branch-cut terms with the dissipative behavior!  In order
to consider the Coulomb interaction we use again the Hubbard
Stratonovich field, $\tilde{a}_\mu = \delta_{\mu.0}\tilde{a}_0$.
\begin{displaymath}
L^{d-wave} \sim \frac{1}{2}(\partial_\mu \alpha - 2
\tilde{a}_\mu)_{\vec{q}, \omega}
\Gamma_2^{(\mu,\nu)}(\vec{q},\omega)(\partial_\nu \alpha - 2
\tilde{a}_\nu)_{- \vec{q},- \omega } - \frac{1}{2}
\rho(\vec{\partial}\alpha)^2 + \frac{1}{e^2 8 \pi}|\vec{q}|^2
\tilde{a}_0 \tilde{a}_0
\end{displaymath}
Now the situation is different, instead $\Gamma_2^{(0,0)}\sim
\rho$ and $\Gamma_2^{(i,i)}\sim 0$ we have; $\Gamma_2^{(0,0)}\sim
\frac{2 \omega^2 - |\vec{q}|^2}{8\sqrt{-\omega^2 + |\vec{q}|^2}}$
and $\Gamma_2^{(i,i)}\sim \frac{|\vec{q}|^2}{8\sqrt{-\omega^2 +
|\vec{q}|^2}}$.  As a result, dimensional analysis shows that the
term $\frac{1}{2e^2}|\vec{q}|^2 \tilde{a}_0 \tilde{a}_0$ is
negligible with respect the term $\Gamma_2^{(0,0)}(q,\omega)
\tilde{a}_0 \tilde{a}_0$.  For this reason we will ignore for the
remaining part the effect of the Coulomb interaction.

Next we present our derivation.  Our starting point is the nearest
neighbor pairing action for a two-dimensional square lattice.  We
start with the superconductor action $\tilde{S}$ and the partition
function $Z$.

\begin{displaymath}
Z = \int D \Phi D\Phi^* D\tilde{C}_\uparrow^\dag
D\tilde{C}_\uparrow D\tilde{C}_\downarrow^\dag
D\tilde{C}_\downarrow \exp(i\tilde{S})
\end{displaymath}
\begin{eqnarray} \nonumber
\tilde{S} &=& \int dt \sum_r \left\{ \sum_{\sigma = \uparrow,
\downarrow} \left[ \tilde{C}_\sigma^\dag(r,t) ( i \partial_t -
a_0^{ext} - E_F) \tilde{C}_\sigma(r,t)\right. \right.
\\ \nonumber &+& \left. t \sum_{\mu = x,y}\left( \tilde{C}_\sigma^\dag
(r+d_\mu, t) \exp^{i\int_{r+d_\mu}^{r} \vec{a}^{ext}d\vec{r}}
\tilde{C}_\sigma (r,t) + h.c.\right) \right]
\\ \nonumber &-&
\sum_{\mu=x,y}\left[ \Phi(r,r+d_\mu; t)
(\tilde{C}_\uparrow^\dag(r,t)\tilde{C}_\downarrow^\dag(r+d_\mu,t)
- \tilde{C}_\downarrow^\dag(r,t) \tilde{C}_\uparrow^\dag
(r+d_\mu,t) ) + h.c. \right]
\\ &+& \left. \frac{|\Phi(r,r+d_\mu; t)|^2}{2
\lambda} \right\}
\end{eqnarray}

The action $\tilde{S}$ is characterized by the nearest neighbor
pairing fields $\Phi$ and $\Phi^\star$ with the coupling constant
$\lambda$ and the external field $a_\mu^{ext}$.  We will perform a
set of transformations which leave the partition function
invariant.  We start by replacing the Fermion field
$\tilde{C}_\sigma(r)$ in terms of chiral fields
$\tilde{R}_{r,\sigma}$ and $\tilde{L}_{r, \sigma}$ ( the right and
left chiral fermions).  $\tau = 1$ corresponds to the nodal liquid
with $K_F = K_{\tau = 1} = ( \frac{\pi}{2 a}, \frac{\pi}{2 a} )$
and $\tau = 2$ for $K_F = K_{\tau = 2} = ( \frac{\pi}{2 a},
-\frac{\pi}{2 a} )$ (see Ref. [5, 6]).  $K_{\tau=1}$ corresponds
to the Fermi surface with the normal in the direction $\hat{e}_1 =
(\hat{x} + \hat{y})/\sqrt{2}$ and $\tau = 2$ is rotated by
$90^\circ$ into the direction $\hat{e}_2 = (\hat{x}-
\hat{y})/\sqrt{2}$.  We replace the Fermion field by
$\tilde{C}_\sigma(r,t) = \sum_{r=1,2} [
\exp(i\vec{K}_\tau\cdot\vec{r}) \tilde{R}_{\tau, \sigma}(r,t) +
\exp(- i\vec{K}_\tau\cdot\vec{r}) \tilde{L}_{\tau, \sigma}(r,t)
]$. Following Ref. [11] we have parametrized the superconductor
order parameter $\Phi$ in terms of the amplitude $\rho(r)$ and the
``Nambu Goldstone'' phase $\alpha(r)$.  This leads to a change of
the measure from $D\Phi D\Phi^\star$ to $D\alpha D\rho$.  We take
the saddle point of the action $\tilde{S}$ and find that the
minimum occurs for $\rho(r,r+d_x) = - \rho(r, r+d_y)$ which
corresponds to the $d_{x^2-y^2}$ symmetry.  We obtain for the
saddle point, $\Delta = 2\sqrt{2} |\rho| = (\frac{2}{\pi})^2
\frac{\hat{\lambda}}{3}$.  In agreement with Ref. [5] we introduce
the spinors $\tilde{\psi}_\tau^\dag =
(\tilde{R}_{\tau,\uparrow}^\dag,
\tilde{L}_{\tau,\downarrow}^\dag)$ and $\tilde{\chi}_\tau^\dag =
(\tilde{R}_{\tau,\uparrow}^\dag, -
\tilde{L}_{\tau,\downarrow}^\dag)$ with $\tau = 1,2$.  In terms of
these spinors we obtain two Dirac Fermion representations with the
Cartesian axes rotated.

Next we perform a gauge transformation [6] which replaces the
spinor fields $\tilde{\psi}_\tau$ and $\tilde{\chi}_\tau$ by the
neutral nodal fermions $\psi_\tau$ and $\chi_\tau$.

\begin{displaymath}
\psi_\tau(r,t)= \exp(-i \frac{\sigma_3}{2}\alpha(r,t))
\tilde{\psi}_\tau(r,t)
\end{displaymath}

\begin{subequations} \label{eq2:whole}
\begin{equation}
\chi_\tau(r,t) = \exp( -i \frac{\sigma_3}{2} \alpha(r,t))
\tilde{\chi}_\tau(r,t) \label{subeq2:1}
\end{equation}
\begin{equation}
\Phi(r,r+d_\mu;t) = \exp(\frac{i}{2}(\alpha r, t)) \rho(r,
r+d_\mu;t) \exp[\frac{i}{2}\alpha(r+d_\mu, t)] \label{subeq2:2}
\end{equation}
\end{subequations}

From eq. 2a we learn that the spinor field is given as a product
of the neutral spinor and the boson field
$\exp(e^{-i\sigma_3}\frac{\alpha}{2}(r,t))$.  In terms of the
fermions fields we have $\tilde{R}_{\tau,\sigma} =
e^{i\frac{\alpha}{2}}R_{\tau, \sigma}$ and $\tilde{L}_{\tau,
\sigma} = e^{i\frac{\alpha}{2}} L_{\tau, \sigma} $ where
$e^{i\alpha/2}$ carries the charge of a half Cooper pair.

As a result of the gauge transformation given by Eq. (2) the
action $\tilde{S}$ is replaced by $\hat{S}$.  We express the
action $\hat{S}$ in terms of the spinors $\psi_\tau$,
$\bar{\psi}_\tau=\psi_\tau^\dag \gamma^0$, $\chi_\tau$, and
$\bar{\chi}_\tau = \chi_\tau^\dag \gamma^0$, where $\gamma^0$ is
the Dirac gamma matrix.  The Dirac action is obtained with the
help of the derivative expansion.  We obtain in the continuum
limit two parts:  The first part is linear in the gauge fields
$(\partial_\mu \alpha - 2 a_\mu^{ext})$ and is given by two Dirac
actions $\hat{S}_0^{(\tau)}$, $\tau=1,2$  (the two directions see
refs 5,6).  The second part is proportional to $(\partial_\mu
\alpha - 2 a_\mu^{ext})^2$ and represents the diamagnetic action,
$S_{dia}^{(\tau)}$.

The Dirac action is characterized by two velocities,
$v_{1,1}=v_{2,2} = 1$ and $v_{1,2}=v_{2,1} = \Delta \equiv
\frac{\tilde{\Delta}}{2\sqrt{2 t}}$ (The first index corresponds
to the nodal liquid $\tau=1,2$ and the second one to the direction
$\mu=1,2$).  We take the expectation value of $\langle
S_{dia}^{(\tau)}\rangle$with respect to the free Fermion and
generate a diamagnetic term [see Eq. (4)].  The diamagnetic mass
in Eq. (4) is given by $r_0(\Delta)\equiv\Lambda\hat{r}_0$,
$\hat{r}_0 = \frac{1}{6 \pi} \int_0^{2\pi} \frac{d\theta}{2\pi}
\frac{\cos^2 \theta}{\sqrt{1+(1-\Delta^2)\sin^2\theta}}$, where
``$\Lambda$'' is the ultraviolet cutoff.  The explicit calculation
of the diamagnetic term is given in ref. 6.  See in particular
eqs. 14, 19, 27 and 28 in ref. 6.

For the remaining part we restrict the calculation to $\Delta \ne
0$ and as a result we obtain two Dirac equations coupled to gauge
fields in 2+1 dimensions: $\hat{S}= \hat{S}_0^{\tau} +
\hat{S}_{dia}^{\tau}$, $\tau = 1,2$,

\begin{equation}
\hat{S}_0^{\tau} = \int d^dx [\bar{\psi}_\tau ( i \tilde{\eth} - m
+ A_\tau) \psi_\tau + \bar{\chi}_\tau (i \tilde{\eth} - m +
A_\tau) \chi_\tau ], \quad m \rightarrow 0
\end{equation}

\begin{equation}
\hat{S}_{dia}^{\tau} = \int d^dx \{ - \frac{r_0(\Delta)}{2}
(A_{\tau,0})^2 \}, \quad d^dx \equiv d^2x dt
\end{equation}
where $A_{\tau,\mu}=\gamma^\mu A_{\tau, \mu}$,
$\tilde{\eth}=v_{\tau,\mu}\gamma^\mu \partial_\mu$, $v_{\tau,\mu}
= (1, v_{\tau,1}, v_{\tau,2}) $, $v_{1,1}=v_{2,2} = 1$,
$v_{1,2}=v_{2,1} = \Delta$.  The form of Eqs. (3) and (4) has been
obtained after performing a derivative expansion in the action in
eq. 1.  $A_{\tau, 0}^2 $ in eq. 4 is the spatial component of the
gauge invariant field $A_\mu$.

Eq. (3) represents the neutral Dirac equation coupled to a gauge
field $A_\tau$ which is a gauge invariant quantity.  Therefore,
the neutral particles $\psi_\tau$ and $ \chi_\tau$ couples to the
gauge invariant field, $ A_0$.  The gauge invariant form follows
from the combination of the Nambu Goldstone phase and the vector
potential $ \vec{a}_\mu^{ext}$.

\begin{eqnarray}  \nonumber
A_{\tau=1,0}&=& a_1 = (a_1^{ext}- \frac{1}{2}\partial_1 \alpha) \\
\nonumber
A_{\tau=2,0}&=& a_2 = (a_2^{ext}- \frac{1}{2}\partial_2 \alpha) \\
\nonumber
A_{\tau=1,1}&=& A_{\tau=2,2} = a_0 = (a_0^{ext}- \frac{1}{2}\partial_t \alpha) \\
 A_{\tau=1,2}&=& A_{\tau=1,2}= 0
\end{eqnarray}

It is important to mention that the gauge field in eq. 3 is
similar to the gauge fields in 1+1 dimensions [9] the reason being
that the second component is zero $A_{\tau=1,2} = A_{\tau=2,1} =
0$, (the index $\tau=1,2$ corresponds to the Fermi surface and the
second index represents the cartesian direction).  Combining the
two gauge fields $A_{\tau,\mu}$, $\tau=1,2$, we obtain a gauge
field in 2+1 dimensions, $A_{\tau=1,0}\equiv a_1$,
$A_{\tau=2,0}\equiv a_2$, $A_{\tau=1,1}=A_{\tau=2,2}\equiv a_0$.
The diamagnetic term in eq. 4 give rise to a Meisner effect,
controlled by the penetration depth, $(r_0(\Delta))^{-1}$
evaluated in eqs. 27, 28 in ref. 6.

Next we will compute the single particle Green's function.  We
will consider first the neutral part.
\begin{eqnarray} \nonumber
G_{\tau= 1}(r,t) &=& \langle\langle T \psi_{\tau=1}(r,t)
\psi_{\tau=1}^\dag(0,0) \rangle\rangle \\ \nonumber &=&
\frac{1}{N} \int D A_{\tau=1} D A_{\tau=2} ( i \gamma^0 \eth +
\gamma^0
A_{\tau=1} )_{r,t; 0,0}^{-1} \quad e^{i \hat{S}_{dia}^{(\tau=1)}} \\
   & & \quad \quad \quad \cdot ( det ( i \tilde{\eth} +
A_{\tau=1} ) det ( i \tilde{\eth} + A_{\tau=2} ) )^2 \quad e^{i
\hat{S}_{dia}^{(\tau=2)}}
\end{eqnarray}

The presence of $\gamma^0$ in eq. 6 is due to the fact that we
compute $\psi_\tau(2) \psi_\tau^\dag(1) =  \psi_\tau(2)
\bar{\psi}_\tau(1) \gamma^0$ and not $\psi_\tau(2)
\bar{\psi}_\tau(1) $.  ``N'' represents the normalization
constant. In order to find the single particle Green's function in
eq. 6 we have to evaluate the determinant; $det(i \tilde{\eth} +
A_\tau) = \exp i \hat{S}_{eff}(A_\tau) $.
\begin{equation}
\hat{S}_{eff}(A_\tau) = \lim_{m \rightarrow 0} i \sum_{n=1}^\infty
\frac{1}{n} \left\{ \verb"Tr" \left[ \frac{i}{i\tilde{\eth} - m} (
- i A_\tau) \right]^n \right\} = \hat{S}_2(A_\tau) +
\hat{S}_I(A_\tau)
\end{equation}

where
\begin{equation}
\hat{S}_2(A_\tau) = \frac{1}{2}\int^\Lambda d^d q A_{\tau, \mu}(q)
\Gamma_2^{\mu\nu}(q) A_{\tau, \nu}(-q)
\end{equation}
with
\begin{equation}
\Gamma_2^{\mu\nu}(q) = 2 \lim_{m \rightarrow 0} \int^\Lambda d^d k
\verb"Tr" \left[ \frac{\gamma^\mu(k + m)\gamma^\nu (k + q +
m)}{(k^2 - m^2)[(k+q)^2 - m^2]} \right] = \frac{1}{8 \sqrt{-q^2}}
( -q^2 g^{\mu \nu} + q^\mu q^{\nu} )
\end{equation}

$\hat{S}_I(A_\tau)$ represents the non-Gaussian part.  In eq. (9)
we used the notation, $g^{\mu\nu}=0$ for $\mu \ne \nu$ and
$g^{00}=g^{ii}=1$.  For the momentum $q$ we use $q^\mu = ( q^0,
\vec{q})$, $q^0 = \omega$ and $q^2 = q^\mu q_\mu = (q^0)^2 -
(\vec{q})^2$.  In eq. (9) we observe that the ``branch-cut'' for
$\omega/|\vec{q}| > 1$ gives rise to an imaginary polarization
[10].  This describes the threshold for destroying the electron
pairs and creating normal electrons.

We combine $\Gamma_2^{\mu\nu}(q)$ given in eq. 7 with the
diamagnetic part $\hat{S}_{dia}^{(\tau)}$ and find the low energy
action in the gauge invariant form.

\begin{equation}
\hat{S}_{2,eff}(A_\mu) = \frac{1}{2}\int^\Lambda d^d q A_{\mu}(q)
\hat{\Gamma}_2^{\mu\nu}(q) A_{\mu}(-q)
\end{equation}

where
\begin{eqnarray} \nonumber
\hat{\Gamma}_2^{(0,0)}(q) &=& \Gamma_2^{(1,1)}(q) +
\Gamma_2^{(2,2)}(q), \quad \bar{\Gamma}_2(q) \equiv
\hat{\Gamma}_2^{(1,1)}(q) = \hat{\Gamma}_2^{(2,2)}(q) =
\Gamma_2^{(0,0)}(q)- r_0(\Delta) \\
\hat{\Gamma}_2^{(0,2)}(q) &=& \Gamma_1^{(2,0)}(q), \quad
\hat{\Gamma}_2^{(0,1)}(q) = \Gamma_2^{(1,0)}(q)
\end{eqnarray}
In eqs. 10, 11 $\hat{\Gamma}_2^{(\mu,\nu)}(q) $ is the vertex for
the fields $a_\mu$ expressed in terms of the original vertex
$\Gamma_2^{(\mu,\nu)}(q)$ (For the fields $A_{\tau,\mu}$see eqs
8,9).

Using eq. 10 we obtain the ``photon'' Green's function.
\begin{eqnarray} \nonumber
D_{00}(\vec{q},\omega) &=& \langle a_0(\vec{q}, \omega)
a_0(-\vec{q}, -\omega) \rangle \\ \nonumber &=& - i \Delta \left[
\frac{\omega^2 - (\Delta q_2)^2}{8(-\omega^2 + q_1^2 + (\Delta
q_2)^2 - i \epsilon)^{1/2}} + \frac{\omega^2 - (\Delta
q_1)^2}{8(-\omega^2 + q_2^2 + (\Delta q_1)^2 - i \epsilon)^{1/2}}
\right]^{-1} \\  &{}_{{ } \overrightarrow{ \Delta \rightarrow 1 {
} }}& \frac{- i \Delta 8 (-\omega^2 + |\vec{q}|^2 -
i\epsilon)^{1/2} }{2
\omega^2 - |\vec{q}|^2 } \\ \nonumber \\
\nonumber D_{11}(\vec{q},\omega) &=&
\langle a_1(\vec{q}, \omega) a_1(-\vec{q}, -\omega) \rangle \\
\nonumber &=&  i q_2^2 \Delta \left[
\left(\frac{q_2}{|\vec{q}|}\right)^2 \left( \frac{q_1^2 + ( \Delta
q_2)^2}{8(-\omega^2 + q_1^2 + (\Delta q_2)^2 - i \epsilon)^{1/2}}
- r_0(\Delta)\right) \right. \\ \nonumber &{ }& \qquad + \left.
\left(\frac{q_1}{|\vec{q}|}\right)^2 \left( \frac{q_2^2 + (\Delta
q_1)^2}{8(-\omega^2 + q_2^2 + (\Delta q_1)^2 - i \epsilon)^{1/2}}
-  r_0(\Delta) \right) \right]^{-1} \\
&{}_{{ } \overrightarrow{ \Delta \rightarrow 1 { } }}& i 8 \Delta
\left(\frac{q_2}{|\vec{q}|} \right)^2 \frac{ (- \omega^2 +
|\vec{q}|^2 - i\epsilon)^{1/2} }{|\vec{q}|^2 - 8  r_0(\Delta)
(-\omega^2 + |\vec{q}|^2 - i \epsilon)^{3/2} } \\ \nonumber \\
\nonumber D_{22}(\vec{q},\omega) &=& \langle a_2(\vec{q}, \omega)
a_2(-\vec{q}, -\omega) \rangle \\ &{}_{{ } \overrightarrow{ \Delta
\rightarrow 1 { } }}&  i \Delta 8 \left(\frac{q_1}{|\vec{q}|}
\right)^2 \frac{ (-\omega^2 + |\vec{q}|^2 - i\epsilon)^{1/2}
}{|\vec{q}|^2 - 8  r_0(\Delta) ( - \omega^2 + |\vec{q}|^2 - i
\epsilon)^{1/2} }
\end{eqnarray}

In the absence of the gauge field $A_\tau$ the neutral fermion
Green's function is given by $G_{\tau=1}^0(\vec{q},\omega) =
\frac{i ( \omega I + \sigma_3 \nu q_1 - \sigma_1 \Delta
q_2}{\omega^2 - (\nu q_1)^2 - (\Delta q_2)^2} $.  Using the
representation in eq. 6 and the dissipative photon Green's
functions given in eqs 12-14, we compute the fermion self energy.
to the one loop approximation we find that the self energy is
given by $\Sigma_\tau(\vec{q},\omega)$.

\begin{equation}
\Sigma_{\tau=1}(\vec{q},\omega) = \omega \int^\Lambda
\frac{d^2k}{(2 \pi)^2} \int \frac{d \Omega}{2\pi}
\frac{1}{(\Omega-\omega)^2-(\vec{k}-\vec{q})^2} \left[
D_{11}(\vec{k},\Omega) + D_{00}(\vec{k}, \Omega)\right]
\end{equation}

The scattering rate in the vicinity of the nodal points,
$|\vec{q}|\approx 0$ is given by the imaginary part of eq. 15. The
main contribution is due to the $D_{00}(\vec{k},\Omega)$ photon.
Performing a contour integral with the pole at $\Omega = \omega +
|\vec{k}-\vec{q}| \sim \omega + |\vec{k}| $, we find in the limit
of $\Delta \rightarrow 1$ the result;

\begin{equation}
Im \Sigma_{\tau=1} (\vec{q}\approx0,\omega) = \frac{\omega^2
\Delta}{ \pi } \int_1^{1+ \Lambda/\omega} dy \frac{(2y -
1)^{1/2}}{1 + y^2} = \frac{4 \omega \Delta}{\pi}
F(\frac{\omega}{\Lambda}); \quad F(\frac{\omega}{\Lambda}) \approx
1 - (\frac{\omega}{\Lambda})^{1/2} + \cdots
\end{equation}

Eq. 16 has been obtained in the limit $\Delta \rightarrow 1$.  The
reason for this being that according to eq. 12 the calculation
becomes simple in this limit.  Eq. 16 represents the ``$\omega$''
dependence of the scattering rate at $T=0$.  This result is much
stronger than the $\omega^3$ result known in the literature [3].

At finite temperatures $T < \Lambda$ eq. 16 gives for the
scattering rate, $\frac{1}{\tau} \sim \frac{T \Delta}{\pi}
F(\frac{T}{\Lambda})$.  Moving away from the nodal point
$|\vec{q}|\simeq0$ gives a reduced scattering rate for
$|\vec{q}|>\omega$, \\
$\frac{1}{\tau} \sim \frac{\omega\Delta}{\pi}
[(\frac{\omega}{|\vec{q}|})^2 - (\frac{\omega}{\Lambda})^2]$.
These results explain the recent ARPES [1] and conductivity [2]
experiments.

Next we consider the Green's function for the quasi-particles,

\begin{equation}
\tilde{G}_\tau(r,t) = \langle\langle T \tilde{\psi}_\tau(r,t)
\tilde{\psi}_\tau^\dag(0) \rangle\rangle \simeq \langle
e^{-\frac{i \sigma_3}{2}(\alpha(r,t)-\alpha(0))} \rangle_A
G_\tau(r,t)
\end{equation}
The phase $\alpha(r,t)$ is given in terms of the gauge field
$a_\mu$ (for $a_\mu^{ext}=0$).

Integration of eqs 5 with the condition $a_\mu^{ext}$ gives for
the Nambu Goldstone phase the result: $\alpha(x,y,t) = 2 [
\int_0^t a_0(0,0;t') dt' + \int_0^x a_1(x',0;t) dx' + \int_0^y
a_2(x,y';t)dy']$.  Using this representation we compute $C(r;t)
\equiv \langle \exp( - \frac{i \sigma_3}{2}(\alpha(\vec{r},t) -
\alpha(0;0)) \rangle_a$.  For simplicity we evaluate the equal
time correlation $C(\vec{R}=R_x, R_y; 0)$,

\begin{eqnarray} \nonumber
C(\vec{R}=R_x, R_y; 0) & \approx & \exp[ - \frac{1}{4} \langle (
\alpha( x + R_x, y + R_y; t) - \alpha(x,y;t)^2 )^2 \rangle ] \\
\nonumber &=& \exp[- \frac{1}{2}\int {d}w \int {d}^2q
[D_{11}(\vec{q},\omega) \frac{1 - \cos q_x R_x}{q_x^2} -
D_{22}(\vec{q},\omega) \frac{1 + \cos q_y R_y}{q_y^2}] \\
&{}_{\overrightarrow{\Delta \rightarrow 1}}& \exp[-\frac{8}{\pi^2}
\hat{r}_0(\Delta) \hat{R}^2] ; \quad \hat{R} =
\sqrt{(\frac{R_x}{a})^2 +(\frac{R_y}{a})^2}
\end{eqnarray}
``$a$'' is the lattice constant and ``$\hat{r}_0(\Delta)$'' is the
dimensionless inverse penetration length. Due to the fact that the
correlation $C(\vec{R})$ is short range allows us to replace the
Green's function $\tilde{G}_\tau$ with the neutral one
$G_\tau(r,t)$, $G_\tau(r,t) \approx \tilde{G}_\tau(r,t)$.  This
replacement is justified for distances $r \leq
\frac{\pi}{\sqrt{2}}(\frac{1}{\hat{r}_0(\Delta)})^{1/2})$ smaller
than the pair correlation function.

Next we want to comment on the effect of the nonlinear term
$\hat{S}_I(A_\tau)$ (given in eq. 7).  Dimensional analysis
suggests that our problem in 2+1 dimensions is equivalent to
Q.E.D. at 3+1 dimensions.  This shift in dimensionality follows
from eqs. 8,9.  To the one loop approximation the nonlinear terms
$\hat{S}_I(A_\tau)$ renormalizes [13,14] the vertex
$\Gamma_2^{\mu\nu}(q)$ (see eq. 3) to $\hat{\Gamma}_2^{\mu\nu}(q)
\sim \frac{1}{8} [ 1 + \frac{14}{3 \pi^2} \ln(\frac{\Lambda}{ -
q^2})]^{1/2} \cdot \Gamma_2^{\mu\nu}(q)$.  Therefore the use of
the normalized vertex $\hat{\Gamma}_2^{\mu\nu}(q)$ will not change
significantly the result in eq. 11.

In the last part we consider the effect of the Coulomb interaction
and show that it can be ignored. Following ref. 12 we replace the
Coulomb term by a Hubbard field $\tilde{a}_0$ with the action
$\frac{1}{2 e^2} \frac{|\vec{q}|^2}{4 \pi}
\tilde{a}_0(\vec{q},\omega) \tilde{a}_0(-\vec{q},-\omega)$.
$\tilde{a}_0$ plays the role of the scalar gauge field.  We
replace in eq. 5, $\vec{a}^{ext}=0$, $a_0^{ext}= \tilde{a}_0$ and
find for the Nambu Goldstone phase ``$\alpha$'' couples to the
scalar field $\tilde{a}_0$ the Lagrangian $L(\alpha, \tilde{a}_0)$

\begin{eqnarray} \nonumber
L(\alpha, \tilde{a}_0) &=& \alpha(\vec{q}, \omega) \left[ \omega^2
\hat{\Gamma}_2^{(0,0)}(\vec{q}, \omega) + |\vec{q}|^2
\bar{\Gamma}_2(\vec{q}, \omega) + \omega q_1
\hat{\Gamma}_2^{(0,1)}(\vec{q}, \omega) + \omega q_2
\hat{\Gamma}_2^{(0,2)}(\vec{q}, \omega) \right] \alpha( - \vec{q},
- \omega ) \\ \nonumber &+& \frac{1}{2}\tilde{a}_0(\vec{q},\omega)
\left[ e^2 \hat{\Gamma}_2^{(0,0)}(\vec{q}, \omega) +
\frac{|\vec{q}|^2}{4 \pi} \right] \tilde{a}_0(-\vec{q},-\omega) \\
&+& \frac{1}{2}\alpha(\vec{q}, \omega) \left[ e ( i \omega)
\hat{\Gamma}_2^{(0,0)}(\vec{q}, \omega)-i q_1
\hat{\Gamma}_2^{(0,1)}(\vec{q}, \omega) - i q_2
\hat{\Gamma}_2^{(0,2)}(\vec{q}, \omega) \right]
\tilde{a}_0(-\vec{q}, -\omega) + h.c.
\end{eqnarray}
The explicit form of the parameters in eq. 19 are obtained from
eq. 11.  In the limit $\Delta \rightarrow 1$ we obtain:
$\hat{\Gamma}_2^{(0,0)}(\vec{q}, \omega) = (2 \omega^2 -
|\vec{q}|^2) \gamma(\vec{q}, \omega)$, $\bar{\Gamma}_2(\vec{q},
\omega) = |\vec{q}|^2 \gamma(\vec{q}, \omega) - r_0(\Delta)$;
$\hat{\Gamma}_2^{(0,i)}(\vec{q},\omega) = \omega q_i
\gamma(\vec{q},\omega)$, $i=1,2$; $\gamma(\vec{q},\omega) =
\frac{1}{8}(-\omega^2 + |\vec{q}|^2)^{- 1/2}$.  From eq. 19 we
observe that contrary to the ``s'' wave case (ref 12) the vertex
$e^2 \hat{\Gamma}^{(0,0)}(\vec{q},\omega) + \frac{|\vec{q}|^2}{4
\pi} {}_{\overrightarrow{q \rightarrow 0}} - \frac{i e^2
\omega}{4}$ not a constant (see ref. 12).  Therefore the
excitations will remain massless in the presence of Coulomb
interactions.

In order to obtain the Nambu-Goldstone excitations we integrate
the field $\tilde{a}_0$ and find:
\begin{eqnarray} \nonumber
L(\alpha) &=& \frac{1}{2} \alpha (q,\omega) \left\{ |\vec{q}|^2
\left[ \omega^2 \left( \frac{ \hat{\Gamma}^{(0,0)}(\vec{q},\omega)
(e^2 4 \pi)^{-1} - \gamma^2(\vec{q},\omega) }
{\hat{\Gamma}^{(0,0)}(\vec{q}, \omega) + \frac{|\vec{q}|^2}{e^2 4
\pi}} + \gamma(\vec{q}, \omega) \right) \right. \right. \\
\nonumber { } & & \qquad \qquad \qquad \qquad - \left. \left.
{r}_0(\Delta) + |\vec{q}|^2 \gamma(\vec{q},\omega) \right]
\right\} \alpha(-\vec{q},- \omega)
\end{eqnarray}
In the limit $|\vec{q}| \rightarrow 0$ we look for massless
solutions $\omega = z |\vec{q}|$.  We substitute $\omega = z
|\vec{q}|$ into the last equation and find a polynomial $P(z)$ for
``$z$''; \\ $P(z) = z^2 (2 z^2 - 1)(1+(e^2 4 \pi)^{-1}(1 -
z^2)^{1/2}) - z^2 - {r}_0(\Delta)(1-z^2)^{1/2}(2 z^2 - 1) = 0$.
Massless solution exists for $z = z(r_0) \approx
\sqrt{\frac{{r}_0(\Delta)}{3(e^2 4 \pi)^{-1} - 1}} $.  We find
finite solutions for $z \approx z(r_0)$ (real and imaginary.),
therefore the excitations are massless in the presence of Coulomb
interactions.

To conclude, a new explanation for the scattering rate is
presented.  Instead of the phenomenological explanation given in
refs 15, 8, we show that the neutral quasi-particles are scattered
by the dissipative collective excitations giving rise to a linear
temperature damping at the nodal points, in agreement with the
ARPES [1] and conductivity [2,3] experiments.  The absence of the
linear scattering rate in the YBCO superconductor might be due to
the absence of the dissipative mode caused by the hopping in the
``c' direction.


\end{document}